\documentstyle[floats,multicol,aps,epsf,prb]{revtex}

\newcommand\relbd{\mathrel{{\bf\smash{{\phantom- \above1pt \phantom-
}}}}}
\newcommand\ltdash{\raise-1.8pt\hbox{$\scriptscriptstyle |$}}
\newcommand \beq  {\begin{equation}}
\newcommand \eeq  {\end{equation}}
\newcommand \bea {\begin{eqnarray} }
\newcommand \eea {\end{eqnarray}}

\newcommand\joinreldex{\mathrel{\mkern-9mu}}
\newcommand\joinrelwx{\mathrel{\mkern-6mu}}
\begin{document}
\draft
\twocolumn[\hsize\textwidth\columnwidth\hsize\csname @twocolumnfalse\endcsname
\title{
Co-operative Two-Channel Kondo Effect
}
\author{ P. Coleman 
$^{*}$
and A. M. Tsvelik}
\address{Department of Physics,
Oxford University
1 Keble Road
Oxford OX2 3NP, UK
}
\author{ N. Andrei  and H. Y. Kee }
\address{
Serin Laboratory, Rutgers University,
P.O. Box 849, Piscataway, NJ 08855-0849, USA.}
\maketitle
\date{\today}
\maketitle
\begin{abstract}

We discuss how the properties of a single-channel Kondo lattice model
are modified by additional screening channels.  Contrary to current
wisdom, additional screening channels appear to constitute a relevant
perturbation which destabilizes the Fermi liquid.  When a heavy Fermi
surface develops, it generates zero modes for Kondo singlets to
fluctuate between screening channels of different symmetry, 
producing a divergent
composite pair susceptibility.  Additional screening channels couple
to these divergent fluctuations, promoting an instability into a state
with long-range composite order.

\end{abstract}
\vskip 0.2 truein
\newpage
\vskip2pc]

A puzzling question that arises in trying to understand heavy fermion
superconductors is how the localized moments seen in high temperature
properties participate in the pair condensate.\cite{heavy}
In these systems a significant fraction of the entropy
associated with the local moments appears to be involved with the
superconducting condensation process: for
$UBe_{13}$, the spin-condensation entropy is about $0.2k_B ln2$ per
spin.\cite{ube13ref} 
The concept of ``composite pairing'', where a Cooper pair and local
moment form a bound-state combination that collectively condenses
may provide a way to understand this large spin-condensation
entropy\cite{emery,balatsky,miranda}  Recent studies
of the one-dimensional Kondo lattice at
strong-coupling\cite{zachar} and the 
infinite dimensional two-channel Kondo lattice have both given
indication of a composite pairing instability.\cite{jarrell}

In this paper we discuss how the properties of a single-channel Kondo
lattice model for heavy fermion systems are modified by 
coupling to additional screening channels.  Current wisdom, based on the
naive
extrapolation from single impurity models,\cite{blandin,coxrev} regards these additional couplings to be
irrelevant.  We shall show that an entirely
different state of affairs arises in a two-channel Kondo lattice where 
the scattering channels of different local symmetry are obliged
to share a single Fermi sea. This 
allows for the possibility
of a {\sl constructive} interference between the two channels
which drives the development of composite order.

Consider 
a sea of conduction electrons coupled to an N-site lattice
of spin-1/2 local moments via two channels:
\bea 
H=H_0+ \sum_j \bigl\{ J_1 \psi{^{\dag}}_{1j} \pmb{$\sigma$} \psi_{1j}
+J_2 \psi{^{\dag}}_{2j} \pmb{$\sigma$} \psi_{2j}\bigr\}\cdot {\bf S_j},
\eea
where
$
H_o=
\sum \epsilon_{\bf k} c{^{\dag}}_{{\bf k}\sigma} c_{{\bf k}\sigma}\label{prim}
$
describes a single electron band and 
\bea
\psi{^{\dag}}_{\Gamma j} =N^{-\frac{1}{2}} \sum_{\bf k} \Phi_{\Gamma {\bf k}} c{^{\dag}}_{\bf k}
e^{-i {\bf k} \cdot {\bf R}_j}, \qquad (\Gamma = 1, \ 2)
\eea
creates an electron at site j in one of two orthogonal
Wannier states,  with form-factor
$\Phi_{\Gamma{\bf k}}$.  
We are motivated to include a weak second-channel coupling
into a Kondo lattice model by the observation that
interactions in the conduction generally 
cause the spin-exchange to spill over from the primary (f-)
channel into a weaker, secondary screening channels.\cite{larkin,tobe}
We shall 
also introduce 
a ``control'' model (II), where 
\bea
H^{(II)}_o = 
\sum_{\bf k \Gamma \sigma} 
\epsilon_{\bf k}
\psi{^{\dag}}_{\Gamma{\bf k}\sigma} 
\psi_{\Gamma{\bf k}\sigma}
\eea
simply describes a band of electrons carrying a conserved
channel quantum number $\Gamma=1,2$.
In the control, electrons in different channels do not mix,
and the absence of a composite pairing
instability in this model 
provides confirmation that that composite pairing effects 
are a consequence of channel interference. 

To examine the effect of second-channel couplings, we introduce the
composite operator \bea \Lambda= \sum_j
-i\psi{^{\dag}}_{1j}\pmb{$\sigma$} \sigma_2 \psi{^{\dag}}_{2j}\cdot
{\bf S_j}, \eea which transfers singlets between channels by
simultaneously adding a triplet and flipping the local moment.  
We now show that channel interference causes 
the susceptibility of this composite operator 
to diverge in Fermi liquid ground-state of channel one.

Suppose $J_2<< J_1$ so that a Kondo effect develops in channel one.
At low energies, the operator
$
(
{\bf S_j} \cdot 
\pmb{$\sigma$}_{\alpha \beta} 
)
\psi_{1 \beta}
$
then behaves as a single bound-state fermion, represented by
the  contraction
\bea
(\mathrel{\mathop{
 {\bf S}_j\cdot
{\pmb{$\sigma$}}_{\alpha \beta} 
) 
\psi_{1\beta}(j) }
^{\displaystyle 
 \ltdash
\joinreldex
\relbd
\joinrelwx\relbd
\joinrelwx\relbd\joinrelwx\relbd
\joinrelwx\relbd
\joinrelwx\relbd
\joinreldex
 \ltdash
{\phantom{_{1\beta}(j)}}
}}
=
{z} f_{j\alpha}.
\eea
where  $z$ is the amplitude for bound-state formation.  
Hybridization between these composite bound-states and 
conduction electrons forms the heavy-fermion quasiparticles,
with energy $E_{\bf k}$ and 
an enlarged Fermi surface whose enclosed volume counts both 
conduction and 
composite f-electrons.\cite{martin,millis,affleck2}

By applying this 
contraction procedure we see that the action of the
composite operator 
$\Lambda$ on the heavy fermion ground-state
{\em creates a pair}:
\bea
\Lambda\vert \Phi\rangle& = &
-i\sum_j
\mathrel{\mathop{ {\bf S}_j\cdot (
\psi^{\dag}_{1j}}
^{
\displaystyle
 \ltdash
\joinreldex
\relbd
\joinrelwx\relbd
\joinrelwx\relbd
\joinreldex
 \ltdash{\phantom{^{\dag}_{1j}}}
}}
\pmb{$\sigma$} \sigma_2 \psi{^{\dag}}_{2j}
)\vert \Phi\rangle \cr
&=&{z}\sum_{{\bf k}, \sigma}
\sigma \psi^{\dag}_{2 {\bf k}\sigma} f{^{\dag}}_{-{\bf k} -\sigma}
\vert \Phi\rangle \eea
In the control model, 
$\psi^{\dag}_{2 \bf k}$  and 
$f^{\dag}_{-\bf k}$ are light and heavy electrons on different
Fermi surfaces.  The mismatch between the decoupled 
Fermi surfaces for channel one and two assures that the excitation
energy $\epsilon_{\bf k }+
E_{\bf k}$ is always finite.  By contrast, in the physical
model, 
$\Lambda$ creates
a pair of heavy quasiparticle  on a {\sl single common} Fermi surface.
To see this explicitly 
we expand both $f_{\bf k}$  and $\psi_{2 \bf k} = \Phi_{2 \bf k}
c_{\bf k}$ 
in terms of quasiparticle operators 
$a_{{\bf k}} 
= \cos \delta_{{\bf k}} c_{{\bf k}} + 
\sin \delta_{{\bf k}} f_{{\bf k}}$.
Near the Fermi surface, scattering is resonant, so 
$\cos\delta_{{\bf k}_F}\sim 1$, whereas
$\sin \delta_{{\bf k}_F}\propto \Phi_{1{\bf k}}$ reflects the
symmetry of the primary screening channel.
Transforming to quasiparticle operators thus
introduces a
factor $\cos ( \delta_{\bf k}) \sin (\delta_{\bf k})
\sim 
\Phi_{1 {\bf k}}$ into the sum, so that near the Fermi surface, 
\bea
\hat \Lambda
\propto\sum _{{\bf k}, \sigma} \sigma \  \Phi_{1  -{\bf k}} \Phi_{2
{\bf k}}  a{^{\dag}}_{{\bf k} \sigma} 
 a{^{\dag}}_{-{\bf k} - \sigma} .\label{pair}
\eea
This relation describes the decomposition of the composite
pair operator in terms of the low-lying quasiparticles.
Notice that the operator takes the form of an
interference between the two channels, and that furthermore,
the two form-factors must
must have the same parity, or the 
composite operator vanishes on the Fermi surface.
Since the excitation energy, $2 E_{\bf k}$ vanishes
on the heavy Fermi surface, it follows that there are now
a large number of zero modes
for the transfer of singlets between
channels. 
\begin{figure}[here]
\epsfysize=1.3in 
\centerline{\epsfbox{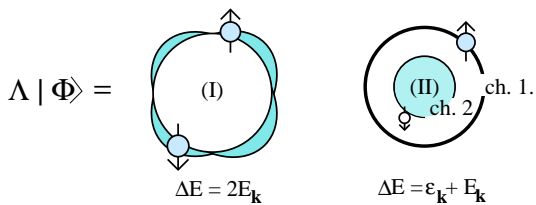}}
\vskip 0.1truein
\protect\caption{Action of composite operator on heavy
Fermi liquid creates: (I) a pair of heavy
fermions (channel interference ) and (II) a heavy and
light electron (channel conservation).
}
\label{Fig1}
\end{figure}

\begin{figure}[tb]
\epsfysize=2.2in 
\centerline{\epsfbox{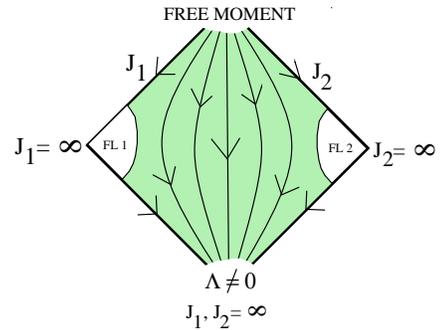}}
\vskip -0.2truein
\protect\caption{
Conjectured renormalization group flows for the co-operative
two-channel Kondo effect. The Fermi liquid formed in channel one or
two is unstable to common a two-channel state with composite order.}
\label{Fig2}
\end{figure}

It follows
that composite pair susceptibility $\chi_{\Lambda}$ must
contain a singular term,
directly proportional to the anisotropic pair
susceptibility of the heavy quasiparticles, 
\bea
\chi_{ \Lambda} \propto \sum_{{\bf k}} \tanh \biggl[
{\beta E_{{\bf k} \lambda}\over 2}
\biggr] {(\Phi_{1 {\bf k}}
\Phi_{2{\bf k}})^2\over 2 E_{{\bf k}}} 
\propto  ln [{T_{K1} \over  T}].
\eea
where $T_{K1}$ is the Kondo temperature for channel one.
Any finite $J_2$ will polarize the transfer
of singlets into
channel two, thereby coupling $J_2$ to this
divergent susceptibility. This will cause $J_2$ to scale
to strong-coupling.
A similar
conclusion will hold when $J_2$ is large, and $J_1$ is small.
Since both Fermi-liquid
fixed points are unstable (Fig. \ref{Fig2}),
continuity of the renormalization flows at strong and weak coupling
leads us 
to  conclude that the physical two-channel
Kondo lattice possesses a new attractive
fixed point which is common to both channels.

One of the distinct 
features of this new lattice fixed point is the development of
off-diagonal composite order,
\bea
\langle \hat \Lambda(x) \rangle \neq 0.
\eea
This type of order involves the explicit
participation of 
two screening channels and the local moments in the pair
condensate.
To explore the nature of this new phase, we present a
simple extension of the existing
mean-field theory of the Kondo lattice. If
we
employ the
pseudo-fermion representation of the local moments
$
{\bf S}_j = \frac{1}{2}f_{j}{^{\dag}}{\pmb{$ \sigma$}}
f_{j}
$,
then the associated constraint
$n_f(j)=1$ at each site leads to a 
local $SU(2)$ symmetry\cite{affleck2}
\bea
f_{j\sigma} 
\rightarrow \left\{
\begin{array}{c}
e^{i \theta} f_{j\sigma},\cr
\cos\phi f_{j\sigma}  + \sigma \sin \phi f{^{\dag}}_{j - \sigma}.
\end{array}\right.
\eea
The Lagrangian for the f-electrons is 
\bea
{\cal L}_f = \sum_{j} \tilde f{^{\dag}} _{ j}
\biggl( \partial_\tau + {\bf W}\cdot {\pmb{ $ \tau$}}\biggr)\tilde f_{j},
\eea
where the tilde field
$
\tilde  f{^{\dag}}_j =
(f{^{\dag}}_{ j\uparrow},\ 
f _{j\downarrow})$ denotes 
a Nambu spinor representation of the f-electron and
${\bf W}$ is a fluctuating gauge field which imposes the constraint.
The  SU(2) symmetry  makes it possible
to simultaneously factorize $H_I$ in both particle-hole and
Cooper channels:\cite{natan}
\bea
H_I
=
\sum_{\Gamma, j} \left\{ \bigl[\tilde f_j{^{\dag}}
{\cal V}^{\Gamma}_j\tilde 
\psi_{\Gamma j}
+{\rm H.c} \bigr]
 + \frac{1}{2J_{\Gamma}}
{\rm Tr}[ {{\cal V}}^{\Gamma\dagger}_j{\cal
V}^{\Gamma}_j]
 \right\}
,
\eea
where 
$\tilde \psi{^{\dag}} _{\Gamma j}= (\psi{^{\dag}} _{\Gamma j \uparrow},
\psi _{\Gamma j \downarrow})$ .
The field
\bea
{{\cal V}}^{\Gamma }_j
= i \left[\matrix{
V& \Delta\cr
-\Delta^*&V^{ *}}
\right]^{\Gamma}_{j},\eea
is directly proportional to  an SU(2) matrix.

The essential observation is that the onsite product of the two fields
${\cal M}(x_j)={\cal V}^{1\dagger}_j{\cal V}^{2}_j$
is invariant under local SU(2)  gauge transformations 
${\cal V}^{\Gamma}_j
\rightarrow
g_j{\cal V}^{\Gamma}_j$. ${\cal M}(x)$ therefore represents
a 
{\sl physical} quantity.
Careful re-expression of this matrix in an operator
form reveals that 
its components 
are directly related to the composite order
that develops between the two channels
\bea
\langle\left[
\matrix{F(x)
&\Lambda(x)
\cr
-\Lambda{^{\dag}}(x)&
F{^{\dag}}(x)
}
\right]\rangle
={{\cal V}^{1 \dagger}(x){\cal V}^2(x)\over J_1J_2}
,
\eea
where
$
F(x_j)= \psi{^{\dag}}_{1j}
{\pmb{ $ \sigma$}}   \psi_{2j}\cdot{\bf S_j} $ represents composite
charge  order and 
$\Lambda(x) $ is the composite pair density.
The product form of this result establishes that
composite order is a consequence
of interference between the Kondo effect in the
two channels.  

By removing the site indices on the hybridization and constraint
field we obtain the mean-field Hamiltonian
\bea
H_{MF} = \sum_{\bf k}
(\tilde c{^{\dag}}_{\bf k},\ 
\tilde f{^{\dag}}_{\bf k} )
\left[
\matrix{
\epsilon_{\bf k}\tau_3 & {\cal V}{^{\dag}}_{\bf k}
\cr
{\cal V}_{\bf k}& {\bf W} \cdot {\pmb{ $ \tau$}}
}
\right]
\left(
\matrix{
\tilde c_{\bf k}\cr
\tilde f_{\bf k} }\right) 
,\label{matrix}
\eea
where the one-band character of the model
forces the order parameter for each
channel to enter into the hybridization
$
{\cal V}_{\bf k}= {\cal V}^1 \Phi_{1{\bf k}}+ {\cal V}^2 
\Phi_{2{\bf k}}
$. 
This provides the origin of the 
interference between the two channels.
Choosing the gauge where ${\cal V}^1= iv_1
\underline{1}$,  then a stable composite-paired
solution emerges with ${\cal V}^2 = v_2\underline{\tau_1}$ and ${\bf W}=
 (0,0,\lambda)$. After some work, we find
that the eigenvalue spectrum of (\ref{matrix})
has two branches, where
\bea
E_{{\bf k} \pm}= \sqrt{
\alpha_{{\bf k}} \pm
(\alpha_{{\bf k}}^2 - \gamma_{\bf k}^2)^{\frac{1}{2}}},
\eea
where
$\alpha_{{\bf k}}= 
 V_{{\bf k} +}^2 + \frac{1}{2}(\lambda^2 +
\epsilon_{\bf k}^2)$, 
$\gamma _{{\bf k}}^2 = [\lambda\epsilon_{\bf k}-V_{{\bf k}-}^2]^2 
+ (2 v_{1{\bf k}}v_{2{\bf k}})^2$
and we have defined
$v_{\Gamma {\bf k}}= v_{\Gamma}\Phi_{\Gamma \bf k}$,
$
V_{{\bf k}{\pm}}^2 = v_{1{\bf k}}^2 \pm
v_{2{\bf k}}^2 
$. 
The requirement that 
the Free energy per site 
\bea
F = - 2 T \sum_{{\bf k}, \alpha = \pm}
{\rm ln} \biggl[
2 \cosh ( \beta E_{{\bf k} \alpha}/2)
\biggr]  + \sum_{\Gamma=1,2}\frac{(v_{\Gamma})^2 }{J_{\Gamma}}.
\label{free}
\eea
is stationary with respect to variations
in $v_2$, $v_1$ and $\lambda$ gives rise to three mean-field
equations.  Two classes of solution exist:

\begin{itemize}

\item
{\bf Normal state}: 
$v_1$ or $v_2= 0$.
Two normal state phases exist
corresponding to a single-channel Kondo
effect in channel one or two. The Fermi surface geometries
of the two phases are topologically distinct, and at half filling
these phases evolve into two different Kondo insulating phases.

\item {\bf Composite paired  state}: 
$v_1v_2>0$. When channel conservation is absent,
a Kondo effect in both channels leads to a paired state
with an anisotropic heavy electron gap function 
$\Delta_{\bf k} \sim \sqrt{T_{K1}T_{K2}} \Phi_{{\bf k}1}
\Phi_{{\bf k}2}$.
\end{itemize}

Setting $v_2= 0^+$ in the mean-field equations, 
the transition from the one-channel
Fermi liquid into the 
composite paired state is given by
$
{J_{2}} \chi_{\Lambda}(T_c)=1
$
where 
\bea \chi_{\Lambda}(T)  = 
\sum_{{\bf k} \alpha} {\rm th}\bigl(
{E_{{\bf k} \alpha}\over 2T}
\bigr)\frac{(\Phi_{{\bf k}2} )^2}{2E_{{\bf k} \alpha}} \biggl[
 1 +  \frac{ (\lambda - \epsilon_{\bf k})^2  }
{(E_{\bf k \alpha}^2 - E_{\bf k -\alpha}^2)
}
\biggr]
\eea
is the composite pair susceptibility.
There are two important contributions to this integral:
a high energy, single-ion part where $E_{\bf k +} \sim \vert\epsilon_{\bf k}\vert >> T_{K1}$ 
and a low energy  ``Fermi surface''
contribution  where the term in square brackets
is proportional to $(\Phi_{1 \bf k})^2$, 
so that
\bea 
\chi_{\Lambda} \approx 2N(0)\biggl[
\langle \Phi_{2{\bf k}}^2 \rangle{\rm ln} \bigl({ D\over T_{K1}}\bigr)
+\langle \Phi_{1{\bf k}}^2\Phi_{2{\bf k}}^2\rangle 
{\mathrm
ln}\bigl({T_{K1}\over T}\bigr)\biggr],
\eea
where $\langle \dots \rangle$ denotes an angular average,
$D$ and $N(0)$  are the conduction electron band-width and density of
states respectively.
Notice how the second interference term
largely compensates for the single-ion
cut-off ($T_{K1}$) in the first term.
A composite pair instability occurs
at 
\bea
T_c \sim D (D/T_{K1})^{\zeta -1}
\exp \left[{-{1 \over 2 \langle \Phi_{1{\bf k}}^2\Phi_{2{\bf k}}^2
\rangle N(0) J_2}}\right].
\eea
where $\zeta = \langle \Phi_{2{\bf k}}^2\rangle/
\langle (\Phi_{1{\bf k}}\Phi_{2{\bf k}})^2\rangle$.

To illustrate this  conclusion we have used
\begin{figure}[bt]
\epsfxsize=2.6in 
\centerline{\epsfbox{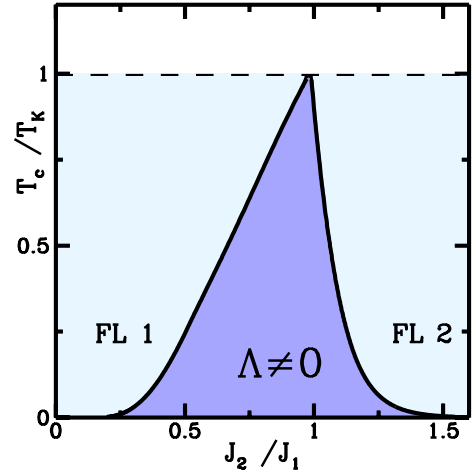}}
\vskip 0.3truein
\protect\caption{Phase diagram
for a two channel Kondo lattice
with ``s'' and ``d-wave'' screening channels.  
Composite pairing develops in shaded region. 
 }
\label{Fig3}
\end{figure}
a two-dimensional model where the local
moments couple to a tight-binding lattice of conduction electrons
via an ``s-'' and  ``d-'' channel:
\bea
\Phi_{1 {\bf k}} = 1, \qquad
\Phi_{2 {\bf k}} = [\cos(k_x)-\cos(k_y)].
\eea
Fig. \ref{Fig3}. shows the phase diagram computed using the mean-field
equations. When $J_2\sim J_1$ the mean-field transition temperature for composite
order is comparable with the single-site Kondo temperature.

\begin{figure}[tb]
\epsfxsize=2.6in 
\centerline{\epsfbox{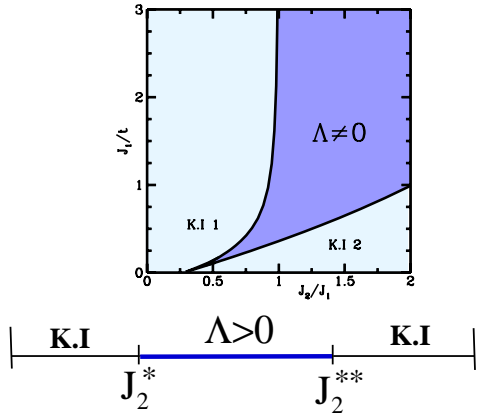}}
\vskip 0.3truein
\protect\caption{Phase diagram for two-channel Kondo insulator. 
``K.I 1'' and ``K.I 2''  denote  Kondo insulating phases
in channel one and two respectively.  In the intermediate gapless
phase both channels participate coherently in the composite pairing
process. }
\label{Fig4}
\end{figure}
An interesting prediction of the theory is the existence
of a second-order superconducting-insulating transition. 
At half-filling the normal
state is a Kondo insulating ground-state
in channel one or two. Beyond a
critical
value $J_2>J^*_{2}$, a Kondo insulator in channel 1
becomes unstable with respect to a composite-paired state.
Even though this phase forms in the complete absence of a Fermi
surface, the superfluid stiffness 
\bea
\rho_s = \biggl({2 \over d}\biggr)\sum_{\bf k} \left({v_1v_2 \over 
V_{\bf k+}}
\right)^2 {
(\Phi_{1\bf k}\pmb{$\nabla$}\Phi_{2\bf k}-\Phi_{2\bf
k}\pmb{$\nabla$}\Phi_{1\bf k})^2\over
[(\epsilon_{\bf k}/2)^2 + V_{\bf k+}^2]^{\frac{1}{2}}}
\eea
is positive (where $d$ is the dimensionality).
At a higher value  $J_2>J^{**}_2$, the Kondo effect in channel one
is finally suppressed, forming 
a second Kondo insulating state.
Fig. \ref{Fig4}. shows how the Kondo-insulating ground-states
become unstable to a composite paired state at strong coupling. 

In closing, it is perhaps instructive to contrast composite and 
magnetically mediated pairing.\cite{miyake,bealmonod}
The latter is maximized
in the  vicinity of
an anti-ferromagnetic quantum-critical point.
By contrast, the composite pairing described here
is driven by a constructive
interference between two rival normal phases, and
requires no fine tuning.
The gap function is determined
by an interference  product of two Wannier functions,
$\Delta_{\bf k} \propto \Phi_{\bf 1 k}\Phi_{\bf 2 k}$, 
predicting an intimate relationship between the gap symmetry
and local quantum chemistry.
When the primary spin exchange occurs in the f-channel, 
a small  exchange coupling to a p-channel will develop a composite paired
state with a gap  symmetry 
$\Phi_{\rm f} \times \Phi_{\rm p}$.
For transition metal systems, 
admixture between a primary d-channel
and a secondary s-channel will provide a gap with d-symmetry.

We should like to thank N. d'Abrumenil, R. Ramazashvili and A. Finkelstein for helpful
comments relating to this work. 
Research was supported in part by the National Science Foundation
under Grants NSF DMR 96-14999 , the EPSRC, UK during a sabbatical stay at
Oxford and NATO grant CRG. 940040. HYK is grateful for support
from a Korea Research Foundation grant. 
\vskip 0.1 truein
\noindent $^*$ On sabbatical leave from Rutgers University.


\vskip 0.1 truein
\begin{references}

\bibitem{heavy}D. L.  Cox  and M. B. Maple, {\em Physics Today} {\bf
48}, 2,32, (1995) and references therein.

\bibitem{ube13ref}J. S. Kim, B.  Andraka and G. Stewart, \prb 44,
6921, (1991).

\bibitem{emery} V. J. Emery and S. A. Kivelson, Phys. Rev. B{\bf 46},
10 812 (1992). 

\bibitem{balatsky}E. Abrahams,  A. Balatsky, D. J. Scalapino 
and J. R. Schrieffer, Phys. Rev. B{\bf 52}, 1271 (1995).

\bibitem{miranda}P. Coleman, E. Miranda and A. Tsvelik, Phys. Rev. B{\bf 49}, 8955 (1994).

\bibitem{zachar} O. Zachar, S. A. Kivelson and V. J. Emery,
Phys. Rev. Lett. {\bf 77}, 1342 (1996); P. Coleman, A. Georges and
A. M. Tsvelik, J. Phys: Cond. Matt {\bf 9}, 345 (1997).

\bibitem{jarrell}M. Jarrell, H. Pang and D. L. Cox, Phys. Rev. Lett.,
{\bf 78 }, 1996 (1997).

\bibitem{blandin} P. Nozi\`eres and A. Blandin, J. Phys. (Paris) {\bf 41}, 193 (1980).

\bibitem{coxrev}D. L. Cox and A. Zawadowski, cond-mat/9704103  (submitted to Adv. Phys.) and
references therein.

\bibitem{larkin}A. I. Larkin and V. I. Melnikov, JETP {\bf 34}, 656 (1972).

\bibitem{tobe}T. Schork and P. Fulde, Phys. Rev. {\bf B 50}, 1345 (1994);
P. Coleman and A. M. Tsvelik, cond-mat/9707003.

\bibitem{martin}R. Martin, Phys. Rev. Lett. {\bf 48}, 362, (1982).

\bibitem{millis}A. Auerbach and K. Levin, Phys. Rev. Lett. {\bf 57},
877, (1986); A. Millis and P. Lee, Phys. Rev. {\bf B35}, 3394, (1986).

\bibitem{affleck2}N. Shibata, K. Ueda, T. Nishimo and C. Ishii,
Phys. Rev. {\bf 54 B},  13495,  (1996).

\bibitem{affleck}I. Affleck, Z. Zou, T. Hsu and P. W. Anderson,
Phys. Rev. {\bf B38}, 745, (1988).

\bibitem{natan}N. Andrei and P. Coleman, J. Phys Cond. Matt {\bf 1} , 4057 (1989).

\bibitem{miyake}K. Miyake, S. Schmitt Rink and C. M. Varma, \prb 34,
6554, (1986).

\bibitem{bealmonod}M. T. B\'eal Monod, C. Bourbonnais \& V.J.  Emery,
\prb 34, 7716, (1986).


\end{references}
\end{document}